\documentclass[12pt]{iopart}
\usepackage{dcolumn}
\usepackage{bm}
\usepackage{graphicx}
\usepackage{color}
\usepackage{subfigure}
\usepackage{hyperref}
\usepackage{latexsym}
\usepackage{amsthm}
\usepackage{amssymb}
\usepackage{cite}
\DeclareGraphicsExtensions{.jpg,.pdf, .mps, .png, .eps, .ps, .EPS,.gif}

\begin{document}
\def\be{\begin{equation}}
\def\ee{\end{equation}}

\def\bc{\begin{center}}
\def\ec{\end{center}}
\def\bea{\begin{eqnarray}}
\def\eea{\end{eqnarray}}
\newcommand{\avg}[1]{\langle{#1}\rangle}
\newcommand{\Avg}[1]{\left\langle{#1}\right\rangle}

\def\ie{\textit{i.e.}}
\def\etal{\textit{et al.}}
\def\m{\vec{m}}
\def\G{\mathcal{G}}

\title{A generalised model for asymptotically-scale-free geographical networks 
}

\author{Nicola Cinardi}
\address{Dipartimento di Fisica e Astronomia {\it Ettore Majorana}, Universit\'a di  Catania, and INFN, via S. Sofia 64, 95123 Catania, Italy}
\ead{cinardinicola@gmail.com}

\author{Andrea Rapisarda}
\address{Dipartimento di Fisica e Astronomia {\it Ettore Majorana}, Universit\'a di  Catania and INFN, via S. Sofia 64, 95123 Catania, Italy\\
Complexity Science Hub Vienna, Josefst\"adter Strasse 39, 1080 Vienna, Austria}
\ead{andrea.rapisarda@ct.infn.it}

\author{Constantino Tsallis}

\address{
Centro Brasileiro de Pesquisas Fisicas and National Institute of Science and Technology for Complex Systems 
\mbox{Rua Xavier Sigaud 150, Rio de Janeiro 22290-180, Brazil} \\
 Santa Fe Institute, 1399 Hyde Park Road, Santa Fe, New Mexico 87501, USA \\
 Complexity Science Hub Vienna, Josefst\"adter Strasse 39, 1080 Vienna, Austria\\}
\ead{tsallis@cbpf.br}
\vspace{10pt}
\begin{indented}
\item[]
\end{indented}

\begin{abstract}
We consider a generalised $d$-dimensional model for asymptotically-scale-free geographical networks.
Central to many networks of this kind, when considering their growth in time,   is the attachment rule, i.e. the probability that a new node is attached to one (or more) preexistent nodes. 
In order to be more realistic, a fitness parameter $\eta_i \in [0,1]$ 
for  each node $i$ of the network is also taken into account to reflect the ability of the nodes to attract new ones. Our $d$-dimensional model   takes into account the geographical distances between nodes,  with different probability distribution for $\eta$ which sensibly modifies the growth dynamics. 
The  preferential attachment rule is assumed to be $\Pi_i\propto k_i \eta_i r_{ij}^{-\alpha_A} $ where $k_i$ is the connectivity of the $i$th pre-existing site and $\alpha_A$ characterizes the importance of the euclidean  distance $r$ for the network growth. For special values of the parameters,
this model recovers respectively the Bianconi-Barab\'{a}si and the Barab\'{a}si-Albert ones.
The present generalised  model is asymptotically scale-free in all cases, and its degree distribution is very  well fitted with $q$-exponential distributions, which optimizes the nonadditive entropy $S_q$, given by 
$p(k) \propto e_q^{-k/\kappa} \equiv 1/[1+(q-1)k/\kappa]^{1/(q-1)}$, with $(q,\kappa)$ depending uniquely only on the ratio $\alpha_A/d$ and the fitness distribution. Hence this model constitutes a realization of asymptotically-scale-free geographical networks within nonextensive statistical mechanics, where $k$ plays the role of energy and $\kappa$ plays the role of temperature. 
General scaling laws are  also found for $q$ as a function of the parameters of the model.

\end{abstract}
\section{Introduction}

Complex networks  constitute a powerful tool for the description of many natural, artificial and social systems. In the last decades hundreds of models have been proposed to describe them \cite{watts98,strogatz01,bara99,newman03,albert,bianconi01,erg02,crucitti04,stefan,deme06,bocca06,bart11,emme14,role2,role,bianconi16,cinardi19}. Of particular importance, and representative of many such systems, are the so-called asymptotically-scale free networks, or simply scale-free networks. In all those models a major role is played by the attachment rule, that is the way in which each element (node) of the system (network) gets new connections (links). The attachment probability for a node to win over the others could be proportional to its degree (i.e., how many connections it already has). If the system is geographically constrained, as for ecological systems, power grids, public transportation, social face-to-face interactions, there will usually be a (inverse) power-law proportionality to the geographical distance between the nodes. In this case the importance of the distance between the nodes can be regulated by the introduction of a parameter $\alpha_A$, where $A$ stands for {\it attachment}. Furthermore, if each node has any ability (or inability) to attract new nodes, a {\it fitness} parameter $\eta \in [0,1]$ can be introduced for every node. 
The possible values that the fitness parameter can take, i.e., the fitness parameter distribution, and the importance of the distances between nodes, open the doors for different models to emerge. 
Among the first ones we find the Barab\'{a}si-Albert model \cite{albert} where there is no dependence on the distances and all the nodes have the same ability to attract new ones (the fitness parameter equals one for all nodes).
A possible extension of this model is the well-known Bianconi-Barab\'{a}si one \cite{bianconi01}, where now the fitness parameter is (introduced and chosen) uniformly between zero and one.\\
It turned out that these are particular cases of more general models \cite{role2,role,stefan}, where the dependence on the distance of the growing mechanism and therefore the role of dimensionality of the system is introduced. The nodes are placed in a specific geographical position (for  dimension $d=1,2$ and 3) based on an isotropic distribution; then the topology of the network is dictated by the degree, the fitness and distances between nodes.\\ 
In this article we generalise the fitness distribution obtaining a new landscape of models of which the above cited are particular cases. We recover them by tuning a new introduced parameter ($\rho$) to particular values as illustrated in the following sections.

\begin{figure}[h]
\centering
\includegraphics[scale=.75,angle=0]{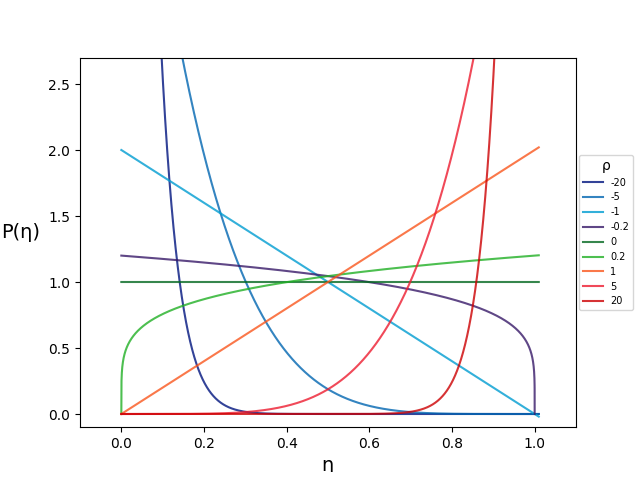}
\caption{\label{fig:Fig1} {Fitness distributions $P(\eta)$ for typical values of $\rho \in (-\infty,\infty)$. The $\rho \to \infty$ and $\rho \to -\infty$ limits correspond respectively to  $P(\eta) =\delta(\eta -1)$ and $P(\eta) =\delta(\eta)$, $\delta(x)$ being the Dirac delta distribution; $\rho =0$ corresponds to the uniform distribution $P(\eta)=1$, $\eta \in [0,1]$.}}
\end{figure}

\section{Model}
In the following we describe the procedure we propose for a generalised asymtotically-scale-free geographical network.
We build the network by successively including one node at a time. We start with the first node placed at the origin, then we add a second node, a third one and so on up to $N$. In all the simulations we run, we have chosen $N=10000$ as the total number of nodes and we run the simulations $1000$ times for each chosen set of parameters (every run took several days). Each node is located at a certain Euclidean distance $r\geq1$ from the center of mass of all the preexisting nodes, and it is picked from a $d$-dimensional {\it isotropic} distribution 
\begin{equation}
p(r)\propto \frac{1}{r^{d+\alpha_G}}
\end{equation}
where $d=1,2,3$, and $\alpha_G>0$ is chosen to make the distribution $p(r)$ normalizable. Here $G$ stands for {\it Growth} to distinguish it from the other parameter introduced here below. As it was shown in \cite{role2}, $\alpha_G$ does not relevantly affect the growth of the network. Therefore we shall typically fix it to be $\alpha_G=2$ in this paper.\\
At each time step, the degree $k$ of each node is updated (after the connections are created as explained here below). 
Also a fitness parameter $\eta \in [0,1]$ is attached to the new arrived node. The main novelty of the present model is the probability distribution of $\eta$ (see Fig. \ref{fig:Fig1}) chosen to be

\begin{equation}
P(\eta)= 
(1+\rho)\eta^\rho, \; \; \; \; \; \; \; \; \; \; \; \; \;  {\it for } \; \; \rho>0  
\end{equation}

\begin{equation}
P(\eta)=1,  \;\;\; \; \; \; \; \; \; \; \; \; \; \; \; \; \; \; \; \; \; \; \; \; \;  {\it for } \; \;  \rho=0 \\
\end{equation}

\begin{equation}
P(\eta)=(1-\rho)(1-\eta)^{-\rho},  \; \; {\it for } \; \; \rho<0 
\label{eq:2}
\end{equation}

\noindent where the pre-factors $(1+\rho)$ and $(1-\rho)$ come from normalization. The just introduced parameter $\rho \in (-\infty,\infty)$ regulates the fitness parameter distribution. The latter can be tuned  to particular values which allows us to recover various well-known models and a variety of new possibilities that we shall discuss along the paper.

When a new node $j$ is added it will be attached to one of the preexisting nodes $i$ following the preferential attachment rule
\begin{equation}
\Pi_i=\frac{ k_i \eta_i r_{ij}^{-\alpha_A}}{\Sigma_i k_i \eta_i r_{ij}^{-\alpha_A}}  \;\;\;(\alpha_A \ge 0)
\end{equation}
where $r_{ij}$ is the geographical distance between nodes $i$ and $j$, and $\alpha_A$ is the parameter that regulates the importance of distances in the attachment rule ($A$ stands for {\it Attachment}). Clearly, in the $\alpha_A=0$ limit, distances play no role and the connectivity is dictated only by how many connections a node already has (i.e. $k$) and its ability to get new ones (i.e. $\eta$).
Basically the topology associated with our model is influenced by the couple of parameters $(\alpha_A,\rho)$, and by  the dimensionality $d$ of the system. \\
Despite its simplicity, the present model is able to reproduce a wide landscape of other models, some of which are well-known,  plus a variety of previously unexplored ones as we will discuss in the following section.

\section{Results}

\begin{figure}[h]
	\includegraphics[scale=.5,angle=0]{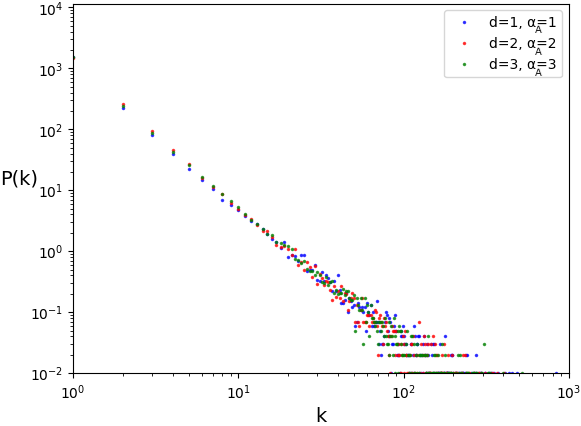}
	\includegraphics[scale=.5,angle=0]{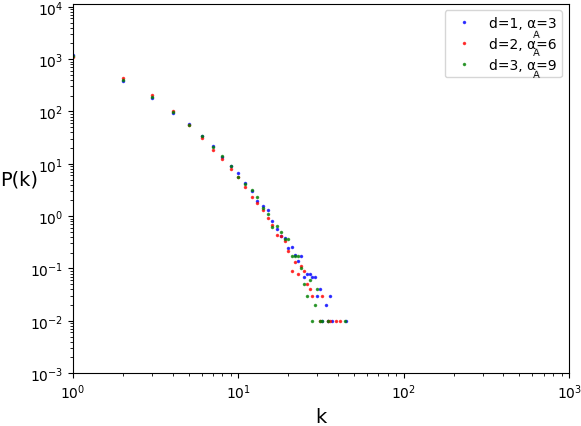}\\
	\includegraphics[scale=.5,angle=0]{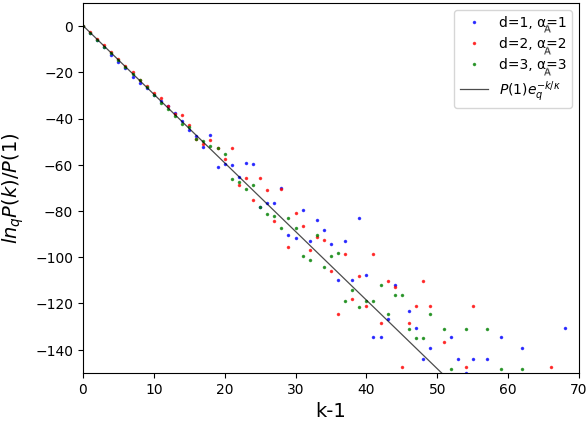}
	\includegraphics[scale=.5,angle=0]{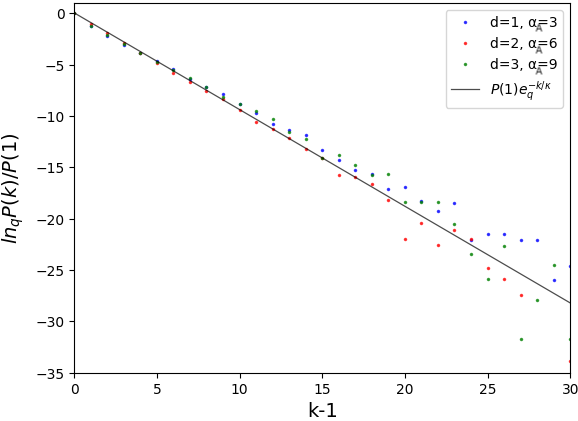}\\
\caption{ {Typical examples of $P(k)$ for $\rho = -100$, as a good approximation for 
$\rho \to  -\infty$ }. {\it Left plots:} For $\alpha_A/d=1$ and $d=1,2,3$ (log--log and $q$-log--linear representations). {\it Right plots:} For $\alpha_A/d=3$ and $d=1,2,3$ (log--log and $q$-log--linear representations).}
\label{fig:Fig2}
\end{figure}

The first natural quantity to study  is the final stationary degree distribution of the network obtained following the growth rules described in the previous section. The latter  is found to be of the form $ P(k) \propto e_q ^{-k/\kappa}$ , see Fig.\ref{fig:Fig2},  where the $q$-exponential function is defined as
\begin{equation}
e_q^z \equiv [1 + (1 -q)z]^{1/(1 - q)} \;\;\;(z \in \rm I\!R)
\label{eq:4}
\end{equation}
if $1+(1-q)z \ge 0$, and vanishes otherwise, with $e_1^z=e^z$.

The distribution $P(k)$ optimizes, under appropriate simple constraints, the nonadditive entropy \cite{tsallis88,tsallis,tsallis09,tsallis19}
\begin{equation}
S_q= \bar k \frac{1 - \sum_i p_i^q}{q-1}  \;\;\;\;\Bigl(\sum_i p_i=1; \, q \in {\rm I\!R}; S_1=S_{BG}=-\bar k \,\sum_ip_i \ln p_i \Bigr),
\end{equation}
where BG stands for {\it Boltzmann-Gibbs}, and $\bar k$ is a positive conventional constant. Since the $q$-exponential distribution optimizes the nonadditive entropy $S_q$, this model constitutes but a particular system within nonextensive statistical mechanics (see \cite{tsallis09,tsallis19} for a review), where $k$ plays the role of energy and $\kappa$ plays the role of temperature. Later on we show how $(q,\kappa)$ depend on $(\alpha_A/d$, $\rho)$. 
As expected we find that the generated networks are asymptotically scale-free.

\begin{figure}[h]
	\includegraphics[scale=.5,angle=0]{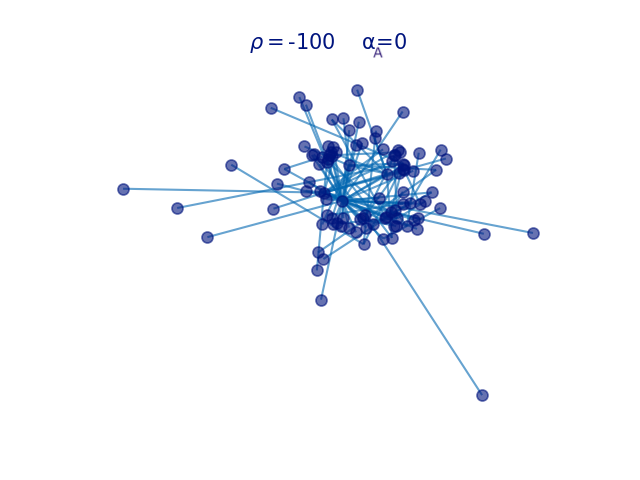}
	\includegraphics[scale=.5,angle=0]{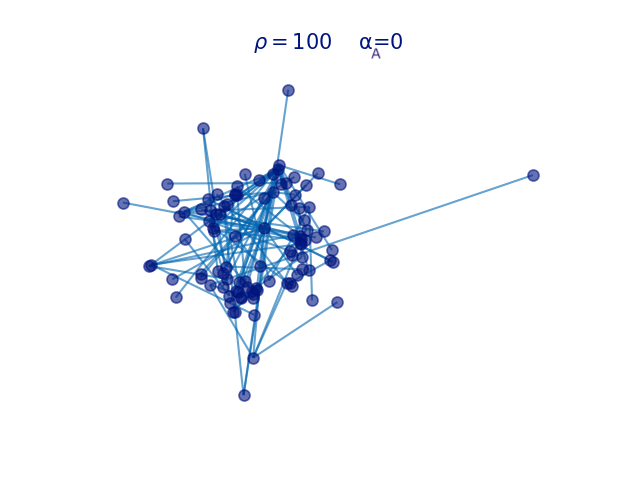}\\
	\includegraphics[scale=.5,angle=0]{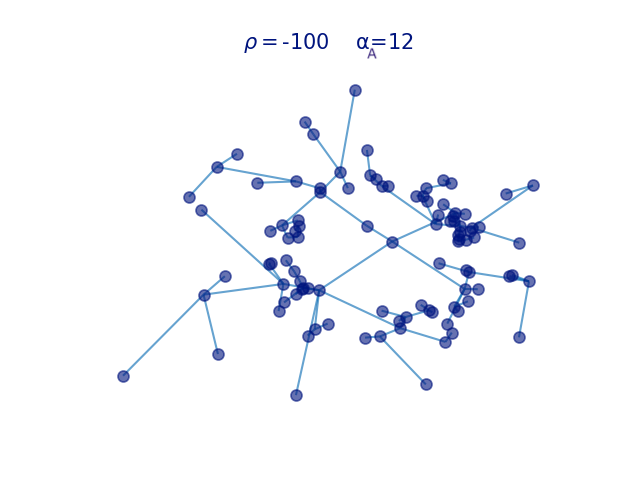}
	\includegraphics[scale=.5,angle=0]{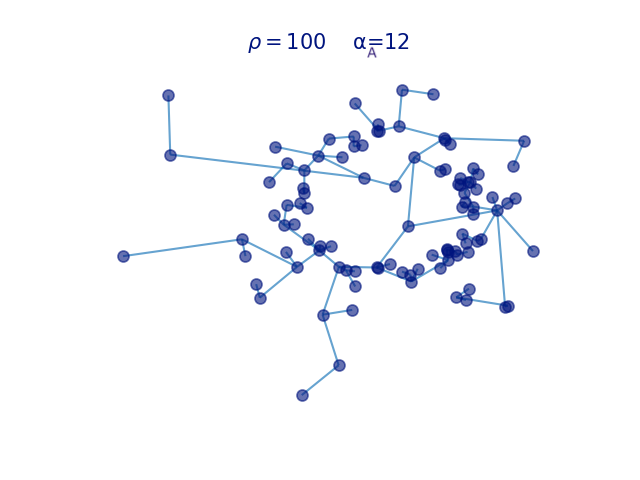}\\
	\caption{\label{fig:Fig3} {$d=2$ stochastic realizations with $N=100$, for typical values of $\rho$ and $\alpha_A$.}    }
\end{figure}

As mentioned above, the model sensibly depends on the couple of parameters $(\alpha_A,\rho)$ plus the dimensionality of the system $d$. 
In particular, for $(\alpha_A,\rho)=(0,0)$ and $(\alpha_A, \rho) =  (0,\infty)$ this model recovers respectively the Bianconi-Barabasi \cite{bianconi01} and the Barabasi-Albert \cite{albert} ones. Indeed, for $\rho=0$ from eq. (3) 
the fitness probability is equal for all nodes and thus the fitness does not play  any role in the model,  while for $\rho \to \infty$ we  get a Dirac-delta function centered at $\eta=1$. The region where $\rho \in [-\infty,0)$ has never been explored before. We built networks with $\rho$ varying in the interval $[-\infty,\infty]$; as expected and shown in fig. \ref{fig:Fig3} (for extreme cases) the parameter $\rho$ does not affect too much the topology of the network while $\alpha_A$ surely does. 
Notice in this figure how the linking and spatial disposition of the nodes are respectively influenced by $\rho$ and by $\alpha_A$.

\begin{figure} [h]
\centering
\includegraphics[scale=.55,angle=0]{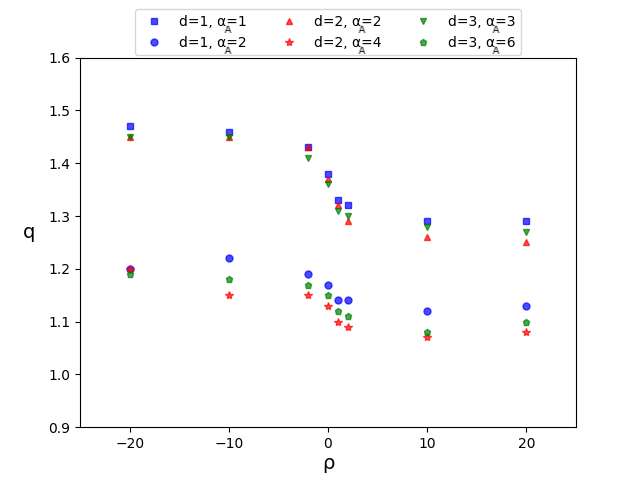}
\includegraphics[scale=.55,angle=0]{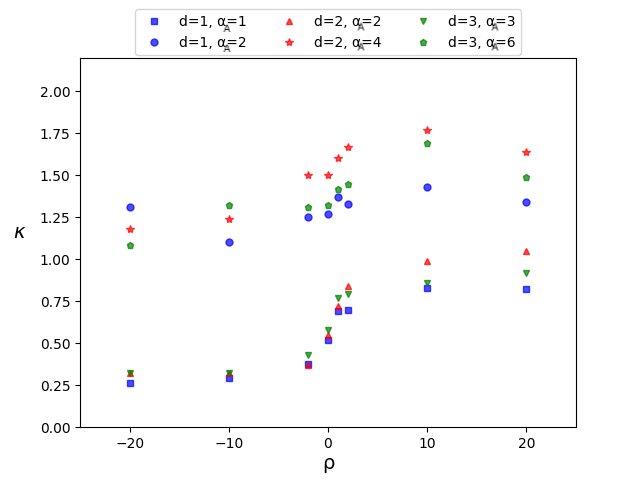}
\includegraphics[scale=.55,angle=0]{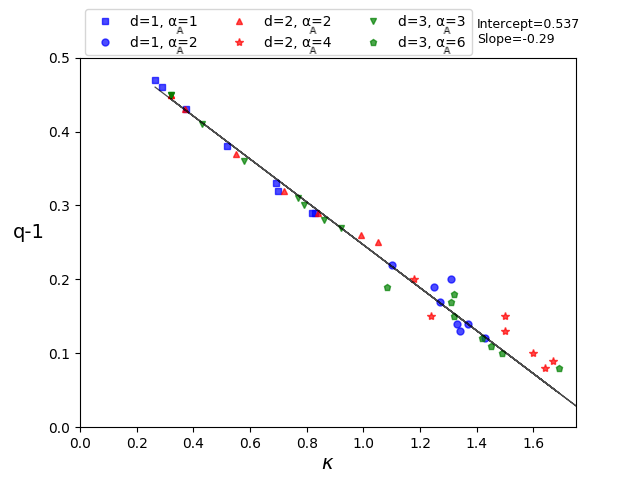}
\caption{\label{fig:Fig4} {{\it Upper plot:} $q$ as a function of $\rho$ for $\alpha_A/d=1,2\;\;(d=1,2,3)$.
{\it Middle plot:} $\kappa$ as a function of $\rho$ for $\alpha_A/d=1,2\;\;(d=1,2,3)$. {\it Lower plot:} Nearly linear relation between $q$ and $\kappa$ (for typical values of $\rho$ and $d=1,2,3$), whose monotonicity denotes the criticality of the system.}} 
\end{figure}

As anticipated, $(q,\kappa)$ depend on $(\alpha_A/d$, $\rho)$. 
Consider first the case $q$ versus $\rho$, see fig.\ref{fig:Fig4}. It is interesting to verify that $q$ varies from a maximum constant value for the part of the spectrum where $\rho \to -\infty$ to a minimum constant value for the part of the spectrum where $\rho \to +\infty$. In the region near $\rho=0$ a more drastic change happens for the values of $q$ (even though all values of $q$ stay in a small interval). In particular $\rho=0$ corresponds to an inflection point for $q$, $\forall \,\alpha_A/d$. Observe the collapse of the curves when $\alpha_A/d=1$ (the upper set of points) or $\alpha_A/d=2$ (the middle set of points) showing the main dependence of $q$ on the ratio $\alpha_A/d$ rather than on $\alpha_A$ and $d$ taken independently. 
The value of $q$ for $0 \le \alpha_A/d \le 1$ numerically approaches 3/2, 7/5 and 4/3 for $\rho$ approaching $-\infty, \,0$ and $\infty$ respectively. Intriguingly enough, these three values of $q$ respectively correspond to the divergences of the moments $\langle k \rangle$, $\langle k^{3/2} \rangle$ and $\langle k^2 \rangle$ of a $q$-exponential distribution: 

\begin{equation}
q=\frac{3}{2} \; \;\; \; \; \;{\it for } \; \; \rho \to -\infty \; \; corresponds \; to <k> \to \infty \\
\end{equation}
\begin{equation}
q=\frac{7}{5} \; \; \; \;\; \;{\it for } \; \; \rho=0 \; \; \;\;\; \; corresponds \;to <k^{\frac{3}{2}}> \to \infty \\
\end{equation}
\begin{equation}
q=\frac{4}{3} \; \;\; \; \; \;{\it for }\; \; \rho \to \infty \; \;\; \;  corresponds\; to <k^2> \to \infty 
\label{eq:6}
\end{equation}

Similar but opposite behaviour for $\kappa$ as a function of $\rho$. More precisely, it goes from a minimum value, when $\rho \to -\infty$ to a maximum value when $\rho \to +\infty$; in the region near $\rho=0$, we observe again an inflection point; the $\kappa$ values change more rapidly than the $q$ values. In any case all points are distributed in a narrow interval. Notice also the same collapse of the curves as for the case of $q$ showing the strict dependence of $\kappa$ on the ratio $\alpha_A/d$ rather than on $\alpha_A$ and $d$ taken independently. An inflection point emerges for both $q$ and $\kappa$ as functions of $\rho$ for $\rho =0$, $\forall \,\alpha_A/d$. The opposite behaviour of $q$ and $\rho$ shows up in the relation that can be observed between the two (see the lowest graph in fig.\ref{fig:Fig4}). All the $(q,\kappa)$ data closely lie within the straight line 
\begin{equation}
q=1.54- 0.29\,\kappa  \quad \forall \,(\alpha_A/d,\rho)
\end{equation}
The monotonicity of this relation constitutes a neat indication of the criticality present in the growing network. Indeed, for each value of $q$, there is only one value of $\kappa$.

\begin{figure}[h]
	\includegraphics[scale=.5,angle=0]{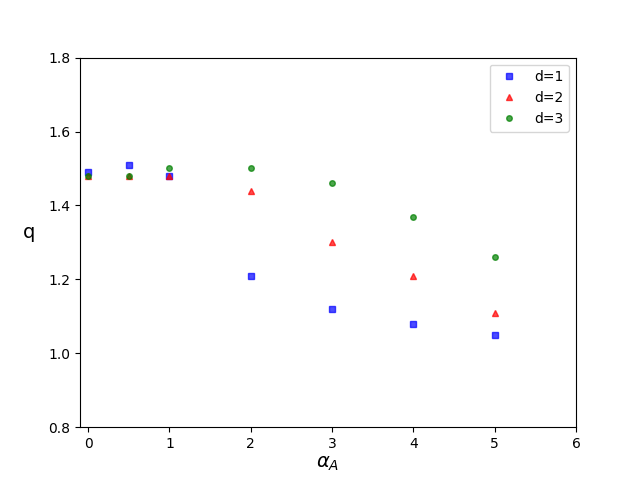}
	\includegraphics[scale=.5,angle=0]{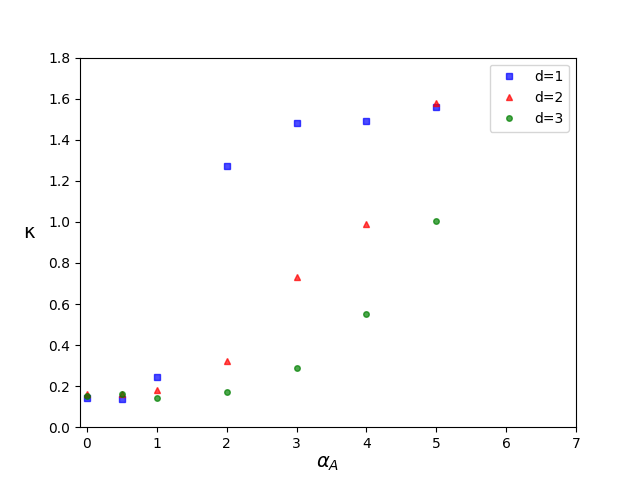}\\
	\includegraphics[scale=.5,angle=0]{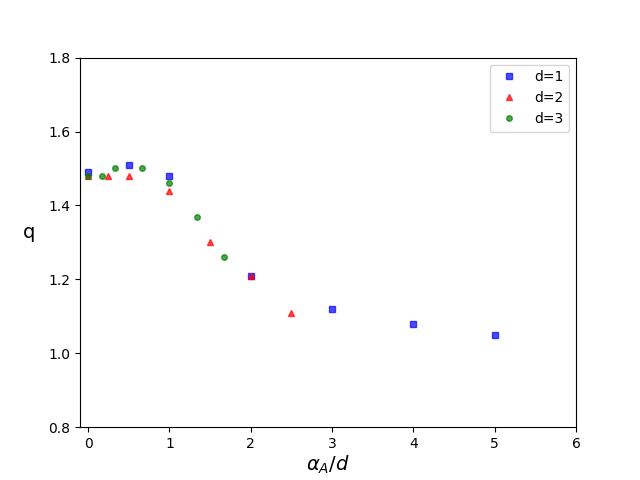}
	\includegraphics[scale=.5,angle=0]{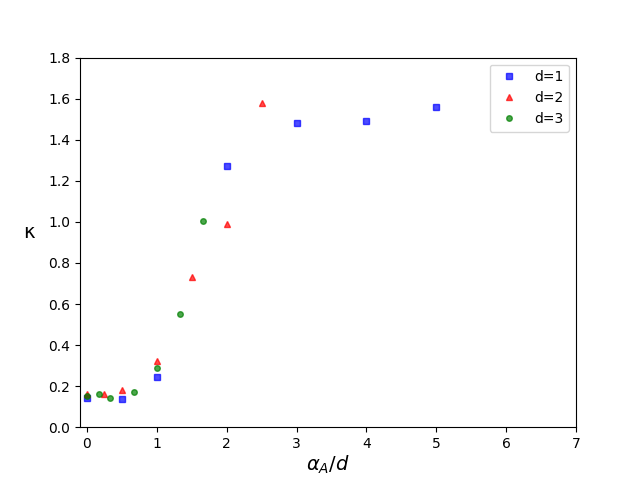}\\
	{\small  
		\caption{\label{fig:Fig 6} {$q$ and $\kappa$ as functions of $\alpha_A$ (upper plots) and of $\alpha_A/d$ (lower plots) for $\rho=-100$ and $d=1,2,3$.}   }}
\end{figure}

\noindent A collapse of the data could be observed also by plotting $q$ and $\kappa$ as functions of $\alpha_A/d$ see fig. \ref{fig:Fig 6}.

We analised different cases, changing the $\rho$ parameter, and we found that the behaviour is very similar for the entire spectrum of $\rho$. Here we show the case for $\rho \to -\infty$, that is an interesting, previously unexplored, case. In the top two plots you can see the dependence of $q$ and $\kappa$, for different dimensions, on the values of $\alpha_A$. They have similar behaviour but differentiate for different dimensions $d$. In the low two plots, it is shown the collapse of the curves when one considers $q$ and $\kappa$ as functions of $\alpha_A/d$. All the curves collapse in a single universal one. As said, this is true for the whole spectrum of $\rho$, showing that our model, with the new introduced parameter is able to reproduce many well-known models and that the universal dependence of $q$ and $\kappa$ on $\alpha_A/d$ is valid for $\rho \in [-\infty, \infty]$, therefore it is a universal behaviour for a pletora of various models. \\

\section{Conclusions}
Summarising, inspired by previous works on asymptotically scale-free network models with fitness parameter $\eta$, geographical constraints, and attachment tuning parameter $\alpha_A$, we have presented a generalised network model with a wider spectrum for the fitness of the nodes. This was achieved by introducing a new probability distribution for the fitness, eqs.(2-4), 
namely by considering the parameter $\rho$ that allows us to fix several different distributions for $\eta$. In particular, by fixing $\rho=\infty$ (and $\alpha_A=0$) we recover the Barab\'{a}si-Albert model, while by fixing $\rho=0$ (and $\alpha_A=0$) it is possible to recover the Bianconi-Barab\'{a}si one. 
The node degree distribution is numerically shown to be $p(k) \propto e_q^{-k/\kappa}$. We have also shown that $q$ and $\kappa$ depend only on $\rho$ and on $\alpha_A/d$. In the $q$ case ($\kappa$ case), as functions of $\rho \in (-\infty,\infty)$, all the values lie in a narrow decreasing (increasing) interval. On the other hand, $\rho=0$ turned out to be an inflection point for both parameters $q$ and $\kappa$, $\forall \,\alpha_A/d$.\\
It was also shown that $q$ and $\kappa$  are dependent on the ratio $\alpha_A/d$ rather than on $\alpha_A$ and on $d$ taken independently. 
Interestingly enough, the value of $q$ for $0 \le \alpha_A/d \le 1$ numerically approaches 3/2, 7/5 and 4/3 for $\rho$ approaching $-\infty, \,0$ and $\infty$ respectively. These values of $q$ respectively correspond to the divergences of the moments $\langle k \rangle$, $\langle k^{3/2} \rangle$ and $\langle k^2 \rangle$ of any $q$-exponential distribution.\\
We have also found, 
 an universal relation between $q$ and $\kappa$ since all the data closely lie along the critical straight line $q=1.56- 0.36\,\kappa$, $\forall \,(\alpha_A/d,\rho)$.\\
Moreover the herein introduced $\rho$ parameter, 
does not affect much the topology of the network, whereas $\alpha_A$ does. This actually is good news for our model in the sense that it constitutes a smooth generalization of the ubiquitous asymptotically-scale-free networks.

\section{Acknowledgements}

N.C. gratefully aknowledges L. Cirto for helpful computational suggestions and S. Brito for several interesting  discussions. He also acknowledges University of Catania and INFN for financial support and Centro Brasileiro de Pesquisas Fisicas in Rio de Janeiro for a very warm hospitality. 
C. T. acknowledges partial financial support of CNPq and Faperj (Brazilian agencies).
 A.R. acknowledges financial support of  the project {\it Linea di intervento 2} of the  Department of Physics and Astronomy {\it Ettore Majorana} of the University of Catania and  of the PRIN 2017WZFTZP {\it Stochastic forecasting in complex systems}.

\end{document}